\documentclass{elsart3-1}

\usepackage{graphicx}

\usepackage{amsmath}
\usepackage{amssymb}

\usepackage[english,francais]{babel}

\newtheorem{e-proposition}[theorem]{Proposition}

\newtheorem{e-definition}[theorem]{Definition\rm}

\renewcommand{\theequation}{\arabic{equation}}
\def\og{\leavevmode\raise.3ex\hbox{$\scriptscriptstyle\langle\!\langle$~}}
\def\fg{\leavevmode\raise.3ex\hbox{~$\!\scriptscriptstyle\,\rangle\!\rangle$}}

\newcommand{\HI}{$\mathrm{H_I}$} 
\newcommand{\Tsys}{T_{sys}}      

\begin{document}
\centerline{The next generation radiotelescopes/ Les radiot\'elescopes du 
futur}
\begin{frontmatter}


\selectlanguage{english}
\title{BAORadio: A digital pipeline for radio interferometry and 21 cm mapping 
of large scale structures}

\selectlanguage{english}
\author[addrLAL]{R\'eza Ansari},
\ead{ansari@lal.in2p3.fr}
\author[addrLAL]{Jean-Eric Campagne},
\ead{campagne@lal.in2p3.fr}
\author[addrGEPI]{Pierre Colom},
\ead{pierre.colom@obspm.fr}
\author[addrIRFU]{Christophe Magneville},
\ead{christophe.magneville@cea.fr}
\author[addrGEPI]{Jean-Michel Martin},
\ead{jean-michel.martin@obspm.fr}
\author[addrLAL]{Marc Moniez},
\ead{moniez@lal.in2p3.fr}
\author[addrIRFU]{James Rich},
\ead{james.rich@cea.fr}
\author[addrIRFU]{Christophe Y\`eche}
\ead{christophe.yeche@cea.fr}

\address[addrLAL]{Universit\'e Paris-Sud, LAL, UMR 8607, F-91898 Orsay Cedex, France 
   \&
    CNRS/IN2P3,  F-91405 Orsay, France }
\address[addrGEPI]{GEPI, UMR 8111, Observatoire de Paris, 61 Ave de l'Observatoire, 75014 Paris, France}
\address[addrIRFU]{CEA, DSM/IRFU, Centre d'Etudes de Saclay, F-91191 Gif-sur-Yvette, France}

\begin{abstract}
3D mapping of matter distribution in the universe through the 21 cm radio emission of atomic hydrogen 
\mbox{(\HI)} is a complementary approach to optical surveys for the study of the Large Scale Structures, 
in particular for measuring the BAO (Baryon Acoustic Oscillation) scale up to redshifts $z \lesssim 3$, 
and therefore constraining dark energy parameters. 
We propose a novel method to map the {\HI} mass distribution in three dimensions in radio, 
without detecting or identifying individual compact sources. This method would require an 
instrument with a large instantaneous bandwidth ($ \gtrsim 100 \mathrm{MHz}$) and high sensitivity, while a 
rather modest angular resolution ($ \sim 10 \, \mathrm{arcmin}$) should be sufficient. 
These requirements can be met by a dense interferometric array or a phased  array (FPA) 
in the focal plane of a large primary reflector, representing a total 
collecting area of a few thousand square meters with few hundred simultaneous beams 
covering a 20 to 100 square degrees field of view. We describe 
the development and qualification of an electronic and data processing system for digital radio interferometry
and beam forming suitable for such instruments with several hundred receiver elements. 


\vskip 0.5\baselineskip

\selectlanguage{francais}
\noindent{\bf R\'esum\'e}
\vskip 0.5\baselineskip
\noindent
{\bf BAORadio:  une cha\^{i}ne num\'erique pour la radio-interf\'erom\'etrie et la cartographie 
des grandes structures \`a 21 cm}
La cartographie de l'univers en radio, \`a travers l'observation de la  raie \`a 21 cm de l'hydrog\`ene atomique, 
constitue une approche compl\'ementaire aux relev\'es optiques pour l'\'etude des grandes  
structures, et en particulier des oscillations acoustiques baryoniques (BAO).
Nous proposons une m\'ethode originale de la mesure de la distribution de l'hydrog\`ene atomique 
neutre \`a travers une cartographie \`a 3 dimensions de l'\'emission du gaz \`a 21 cm, sans 
identification de sources compactes. Cette m\'ethode n\'ecessite un 
instrument avec une grande sensibilit\'e et une grande largeur de bande instantan\'ee 
($\gtrsim$ 100 MHz), alors qu'elle peut se contenter d'une r\'esolution angulaire moyenne (10 arcmin). 
Ces performances peuvent \^{e}tre obtenues avec un r\'eseau  interf\'erom\'etrique compact
ou un r\'eseau phas\'e au foyer d'un grand r\'eflecteur, repr\'esentant une surface 
collectrice de quelques milliers de m$^2$ avec quelques centaines de lobes simultan\'es couvrant 
un champ de vue de 20 \`a 100 deg$^2$. Nous d\'ecrivons le d\'eveloppement et la qualification 
d'un syst\`eme \'electronique et de traitement pour la radio-interf\'erom\'etrie et la synth\`ese de 
lobe num\'erique, adapt\'e \`a ce type d'instrument avec 
plusieurs centaines d'\'el\'ements de r\'eception.   

 \vskip 0.5\baselineskip
\noindent{\small{\it Mots-cl\'es~:} Cosmologie; \'Energie Noire; Radio-interf\'erom\'etrie \`a 21 cm; }}
\end{abstract}
\end{frontmatter}

\selectlanguage{english}
\section{Introduction}
\label{sec:intro}
The nature and properties of Dark Energy are among the most puzzling questions in physics 
and cosmology today. Dark Energy has been hypothesized as a mysterious energy density component 
present in the Universe responsible for its accelerated expansion \cite{amend.10,copeland.06}. 
The acoustic oscillations of the photon-baryon plasma prior to the decoupling are the source of modulations 
or wiggles present in the spatial power spectrum of matter distribution in the universe, usually referred to as BAO 
(Baryon Acoustic Oscillations). These acoustic oscillations have clearly been observed in the anisotropies 
of the Cosmic Microwave Background \cite{wmap.11}.
The measurement of the oscillation wavelength at different redshifts is a standard ruler type cosmological 
probe and is considered as one of the most robust probes of the Dark Energy equation of state \cite{DETF}.

The matter distribution in the Universe is often observed in the optical window using galaxies as tracers. 
The BAO scale has indeed been already measured by optical photometric or spectroscopic 
surveys up to redshifts $z \lesssim 0.7$ (see e.g. \cite{percival.10}). 

Another possible tracer is the neutral atomic hydrogen ({\HI}) gas which can be detected through 
its 21 cm emission or absorption. Radio surveys to map matter
distribution using {\HI} 21-cm emission 
are complementary to optical surveys to study the Large Scale Structure (LSS) statistical properties. 
A novel method to perform such a survey through intensity mapping has been 
suggested by  J.B. Peterson and collaborators \cite{peterson.06}. It should be stressed that 
the observation method discussed here {\it (Intensity Mapping)} 
aims at the measurement of the total 21 cm radiation intensity as a function of the redshift $z$, 
without detection of individual compact radio sources (galaxies \ldots) \cite{chang.08,ansari.08,seo.10}.
Such a survey would produce a 3D map of {\HI} emission (2 angles + frequency $\sim$ redshift), with an 
angular resolution of 10-20 arc minutes, and $\sim$ 50 kHz in frequency ($d z/z < 10^{-3}$). 

A research and development program (BAORadio)  in electronics for radio interferometry for the GHz frequency 
domain has been initiated by LAL (CNRS/IN2P3)\footnote{\tt http://www.lal.in2p3.fr} and 
Irfu (CEA)\footnote{\tt http://irfu.cea.fr} at the end of 2006.
Observatoire de Paris\footnote{\tt http://www.obspm.fr} 
(GEPI and Nan\c{c}ay radio observatory) is also a major partner of this program since the summer 2007. 
A complete analog and digital electronic pipeline, with the corresponding firmwares and acquisition and processing 
softwares has been designed and developed, by a team with broad technical and scientific expertise from 
partner laboratories and the financial support from P2I\footnote{GIS Physique des Deux Infinis}, 
Irfu (CEA), IN2P3 (CNRS) and PNCG\footnote{Programme National Cosmologie et Galaxies}.
This system has been used in interferometric mode to characterize the cylindrical reflectors 
built by our colleagues at the University of Carnegie-Mellon (CMU) at Pittsburgh \cite{bandura.11}.
The FAN project (Focal Array at Nan\c{c}ay) uses also the electronic and acquisition system developed 
for the BAORadio project. A phased array multi-beam receiver for the Nan\c{c}ay Radio Telescope (NRT) has been 
developed and is being characterized at the focal plane of NRT in the framework of the FAN project. 
This can lead to the development of a multi-beam system which can enhance the sensitivity of the NRT by an 
order of magnitude in survey mode \cite{martin.11}. 

The scientific issues and specific difficulties of the observation
method proposed here, for the 3D mapping of the 
21 cm emission temperature as a function of frequency {\it (Intensity Mapping)}, will be presented in section 2. 
We will more specifically discuss the sensitivity needed to measure the spatial power spectrum 
of matter density inhomogeneities $P(k)$ and the scientific and technical challenge of separating the 
cosmological signal from the radio foregrounds. The electronic and processing system 
developed for the BAORadio project will be presented in section 3 and some preliminary results 
obtained during different qualification tests at Nan\c{c}ay and Pittsburgh will be briefly discussed in section 4. 

\section{3D mapping of the 21 cm emission}
The statistical properties of the matter distribution in the universe or the Large Scale Structures (LSS) 
can be used to test the cosmological model and to determine the values of the cosmological parameters, 
such as the matter or dark energy densities. 
We can observe the cosmological matter distribution, dominated by dark matter, 
through tracers like galaxies in most cases. These galaxies are usually detected and observed through their 
optical emissions, the position being determined by imaging and the redshift by spectroscopy. 
A similar method can be used for galaxies and compact gas cloud in radio, through the 21 cm  
hyperfine transition (1.42 GHz at $z=0$) of neutral atomic
hydrogen. Here, the source redshift $z$ can be 
directly determined by comparing the observed radio frequency $\nu_r$ to the intrinsic emission 
frequency ($z = \frac{1420 \mathrm{MHz}}{\nu_r} - 1$). A discussion of
the cosmological applications of observation 
of galaxies at 21 cm can be found for example in references \cite{abdalla.05,ska.science}. 

The relatively low radio brightness of these objects limits their detection to the vicinity, in cosmological 
sense, of our galaxy with the current instruments (e.g. \cite{alfalfa.10}). 
An instrument with a collecting area $A \sim 10^4 \, \mathrm{m^2}$
like the Large Nan\c{c}ay Radio telescope (NRT) can only detect galaxies with $M_{H_I} \sim 10^{10} M_\odot$ 
in 21 cm up to redshift $z \lesssim 1$ with an integration time of few hours ($S_{lim} \sim 1 \, \mathrm{mJy}$).
Radio detection of galaxies at 21 cm would thus need collecting areas  $A \sim 1 \, \mathrm{km^2} = 10^6 \, \mathrm{m^2}$, 
as envisaged for the final phases of the SKA project\footnote{\tt http://www.ska.org}. 

The power spectrum of density inhomogeneities $P(k)$ is one of the most widely used mathematical tools to quantify 
and analyse statistical properties of matter distribution in cosmology.  If $\rho(\vec{r})$ denotes matter density  
in the cosmos and $F(\vec{k})$ the Fourier transform of the matter density inhomogeneity field $\delta(\vec{r})$, we have:
$$ \delta(\vec{r}) = \frac{\rho(\vec{r})}{\bar{\rho}} -1 \hspace{2mm}  \rightarrow TF \rightarrow \hspace{2mm} F(\vec{k})$$ 
Assuming isotropy at large scales, the statistical properties of the matter distribution can be represented 
by the power spectrum $P(k)$. 
$$ P(k) = < | F(\vec{k}) |^2 > \hspace{3mm} \mathrm{avec} \hspace{3mm} k = |\vec{k}| $$
Interested readers might refer to  \cite{cosmo.rich} or \cite{cosmo.peacock} for in depth presentation of 
the cosmological model, structure formation and matter power spectrum.

Baryon Acoustic Oscillations (BAO) correspond to inprints left over by the matter distribution at large scales
by the acoustic (pressure) waves which were propagating in the photon-baryon plasma prior to decoupling. 
The BAO appear as a slight modulation ($\lesssim 10 \%$ in amplitude)  of the power spectrum 
$P(k)$ for $k_n \sim 2 \pi \, n \, (1/150 \mathrm{Mpc})$. This length scale (150 Mpc) corresponds to 
transverse structures with an angular scale around a degree and to
longitudinal structures with $\delta z\sim 0.045$ at $z=0.5$, $\delta z\sim .06$
at $z=1.0$ and $\delta z \sim 0.1$ at $z=2.0$.
The mean brightness temperature at 21 cm 
and the corresponding inhomogeneities level are smaller than a milli-kelvin \cite{barkana.07,ansari.11}. 
A radio instrument with a rather modest size  ($D \lesssim 100 \, \mathrm{m}$) should thus be able 
to resolve these structures. However, a high sensitivity to brightness temperature variations 
is required to avoid the signal to be swamped by the instrumental noise, quantified by 
the system temperature ($\Tsys$). The determination of the power spectrum indeed requires
to survey a large volume of the universe in order to limit fluctuations due to the limited 
number of measured Fourier modes {\it (Sample Variance)}. One has to cover a significant fraction 
of the sky ($\sim 5000-10000 \, \mathrm{deg}^2$)  and a sufficiently thick slice in redshift $\Delta z \gtrsim 0.2$
to maintain the sample variance under control.

\begin{figure}
\centering
\mbox{
\vspace*{-10mm}
\includegraphics[width=0.8\textwidth]{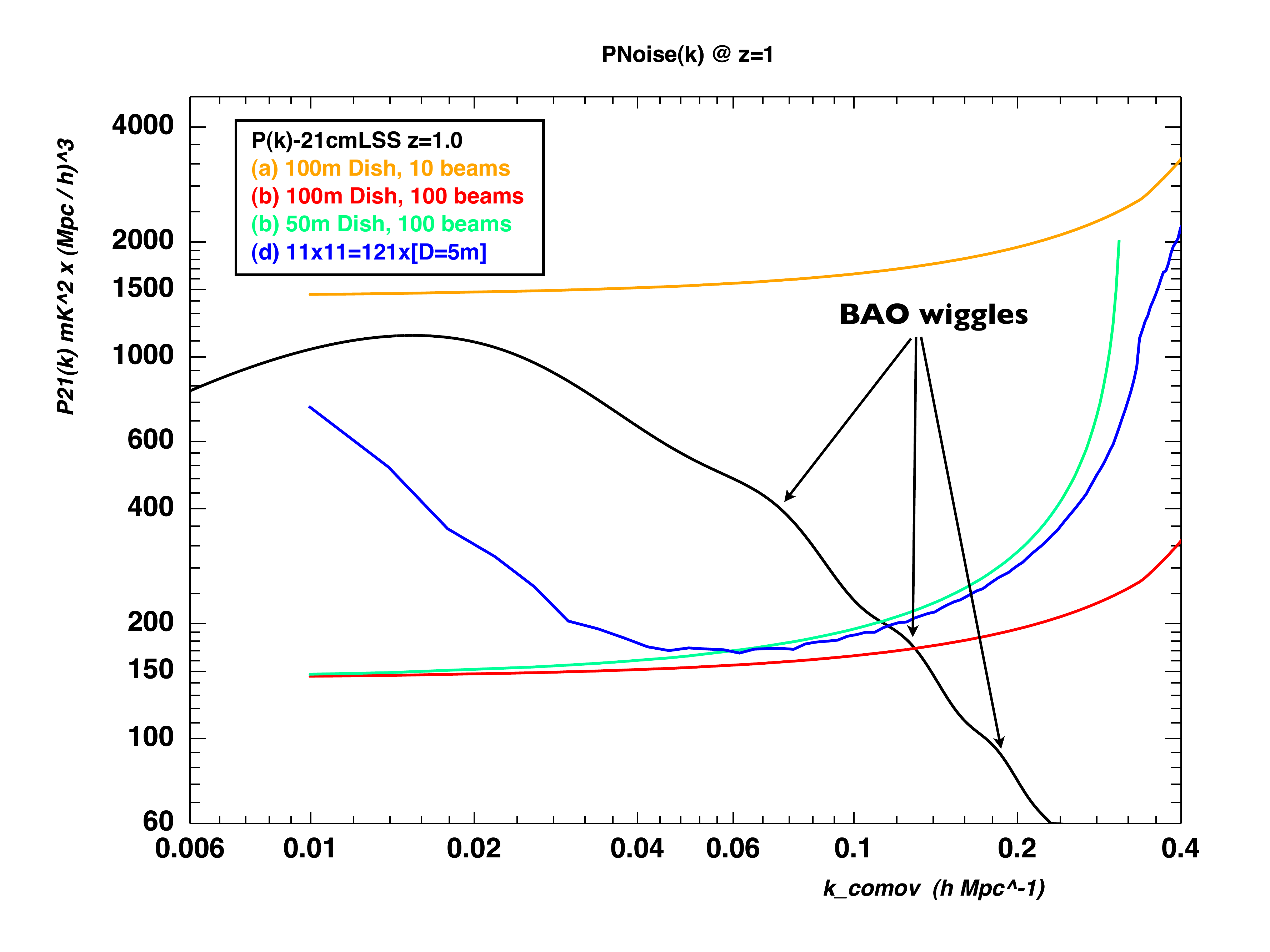}
}
\vspace*{-5mm}
\caption{ Noise power spectrum $P_{noise}(k)$ of a radio instrument compared to the 
 expected 21 cm brightness temperature power spectrum $P_{21}(k)$. We have assumed the survey of a quarter 
of the sky ($\pi$ srad) over a one year period and $\Tsys = 50 \, \mathrm{K}$ system temperature. 
The noise power spectrum is shown for four different instrument configurations:
(a) D=100 meter diameter reflector, equipped with 10 beam FPA (Focal Plane Array), 
(b) D=100 meter diameter reflector and 100 beams, (c) D=50 meter diameter reflector and 100 beams, 
(d) a dense interferometric array with $11 \times 11 = 121$ D=5 meter diameter dishes, covering 
a $55 \times 55 \mathrm{m^2}$ area.  }
\label{figsenspk}
\end{figure}

These constraints can be fulfilled by a wide band ($\gtrsim 100-200 \mathrm{MHz}$) radio instrument 
with a large instantaneous field of view  (20-100 $\mathrm{deg^2}$), which will provide
three dimensional (2 angles and frequency) maps of the 21 cm brightness temperature 
with $\sim \mathrm{10 \times 10 \, arcmin^2}$ angular resolution and 
$\delta \nu \sim \mathrm{50-100 \, kHz}$ frequency resolution. These maps could then 
be used to determine the atomic hydrogen spatial power spectrum. 
Figure \ref{figsenspk} shows the contribution of instrumental noise for various 
instrument configurations for a radio survey covering a quarter of the sky (10000 deg$^2$)
over one year. The power spectrum of the expected spatial variations
of the 21 cm brightness temperature is shown, as well 
as the BAO wiggles\cite{ansari.11}.    

\begin{figure}
\centering
\mbox{
\vspace*{-10mm}
\includegraphics[width=\textwidth]{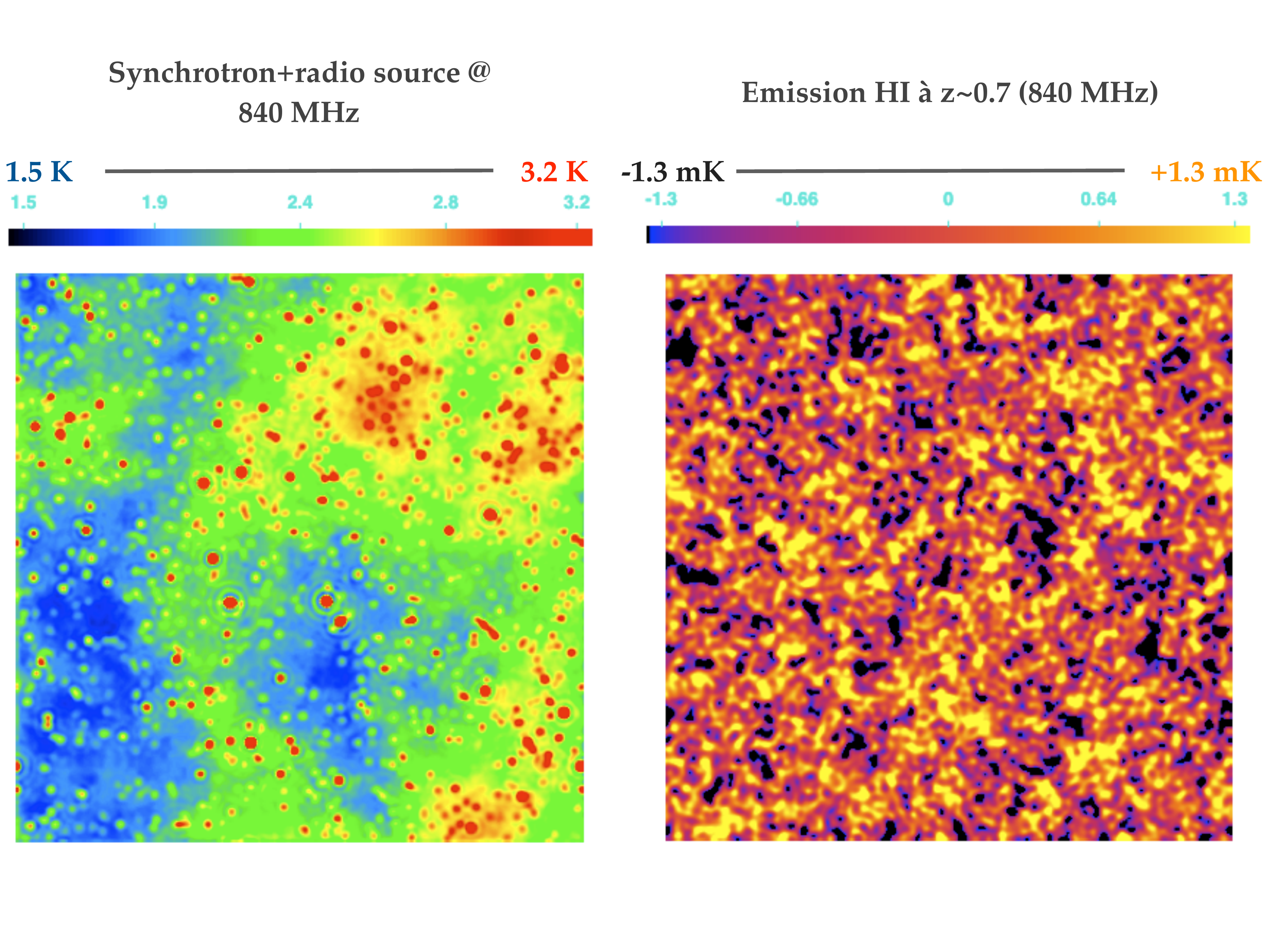}
}
\vspace*{-20mm}
\caption{Left: Brightness temperature of Galactic synchrotron and continuum radio sources 
at 840 MHz (color scale in Kelvin) for one of the $30 \times 30$ deg$^2$ coldest patches of the sky, 
Right: Brightness temperature expected for the 21 cm emission of atomic hydrogen at z=0.7 
(color scale in milli-kelvin)  \cite{ansari.11} }
\label{figradsrc}
\end{figure}

The level of instrumental noise is not the only experimental challenge of the {\it Intensity Mapping} method. 
Indeed, the radio foregrounds due to the synchrotron emission of the
Milky Way and the continuum 
emission of compact radio sources are a thousand times brighter than
the {\HI} emission.
Figure \ref{figradsrc} provides a simulated map showing the expected level of variations of the 
21 cm emission of neutral atomic hydrogen at $z \sim 0.7 \, (\nu \simeq 840 \mathrm{MHz})$, 
compared to the Galactic synchrotron and radio source temperature, for
one of the coldest regions of the 
sky. To fight against this overwhelming sky emission level, 
several authors have suggested to use the spectral power law behaviour 
of the synchrotron emission $\propto \nu^\beta$ 
to separate the cosmological signal from the foreground and background radio 
contamination. Encouraging results have been obtained with preliminary
studies using such a method for component separation, based on
simulations \cite{ansari.11}.    

\section{BAORadio electronic pipeline}
\label{secdevelec}
End of 2006, following J.Peterson and U-L Pen proposal for a 21 cm intensity mapping 
survey for dark energy, LAL \& Irfu initiated an ambitious R\&D effort for the development of an electronic 
system suitable for interferometric radio surveys in UHF/L bands. Our engineering team has designed 
and built a complete analog and digital electronic system for acquisition and processing of radio signals.

The striking feature of the system resides in its nearly fully digital design, the use of FPGA based,
application oriented custom designed boards, to obtain large bandwidth digital processing, while 
keeping low power consumption. In addition, the system has been designed to be scalable to large number 
of channels (few hundreds to thousands analog channels), and distributable over a large area (few hundred meters).
Other digital system, with similar features for radio astronomy \cite{dsp.radio}, have been developed, or are 
under development: CASPER/ROACH at Berkeley \cite{casper}   and Uniboard in Europe \cite{uniboard}. 

Analog signals are digitized at 500 Msample/s, 
providing a 250 MHz total instantaneous bandwidth in the current system version. The digital data flow is then 
processed through the different stages, either by firmware in FPGA's or by software in the acquisition computer 
cluster. The system has been mainly designed and developed during the three years 2007-2009. In addition to 
measurements using test bench in the laboratory, the system has undergone extensive qualification tests 
at Nan\c{c}ay using the large radio telescope (NRT), or with the cylindrical reflector prototypes at Pittsburgh.
The system is composed of several components which are briefly described below. A more technical description 
of the system and its performance can be found in \cite{charlet.11}.

\begin{enumerate}
\item The analog modules {\bf AEM} perform  two main functions : amplification, filtering of the 
RF signal, and frequency down conversion to intermediate frequency (IF). The current boards have filters
designed to mix the 1250-1500 MHz band down to 0-250 MHz,
with the local oscillator at 1250 MHz.
\item The clock and control signal distribution {\bf DCLK} is a crucial subsystem. 
DCLK is designed to transfer a master clock from a central facility 
to slave clock and control signal synthetizer boards. These secondary boards, 
located close to the reflectors, distribute the master clock to the analog modules and 
the sampling (digitizer) boards. 
This is a key component for building a large digital interferometer.
\item The {\bf DIGFFT} sampling board features 4 analog inputs which are sampled at 500 
Msample/s with 8 bits dynamic range. The board is equipped with powerful digital circuits
 (FPGA) capable of processing on the fly the digital stream to separate the signal into frequency 
components, through an FFT like digital filter in firmware. 
The data are transmitted through two high speed (5 Gbits/s) optical fibers to the acquisition 
computers, or to a dedicated beam-former, located in a central building. 
These boards are equipped with USB and VME interfaces used to configure the circuits and for slow control.
We have developed a set of firmwares for the DIGFFT boards: the RAW firmware 
samples and transfers over optical fibers the corresponding raw data with 
a maximum of 48 kSample per digitization frame; the FFT firmware performs sampling, Fourier 
transform and transfer of the Fourier coefficients (2 bytes, real/imaginary). The FFT is performed 
on 8 kSample data chunks per channel, yielding a frequency resolution of $\sim$ 61 kHz. 
The 8 bit dynamic range of the RAW data and truncated FFT coefficients should give enough dynamic range to 
identify and clean RFI (Radio Frequency Interference) due to terrestrial radio sources.
\item A set of PCI-Express optical data receivers {\bf PDR} modules receive the data from the 
optical stream and transfer it to the acquisition system memory. We have developed the firmware 
for the PDR boards, as well as the acquisition software which uses the Jungo 64 bit low level PCI-Express driver.
Data is exchanged between DIGFFT and PDR boards, as well as between different computing nodes in a
cluster embedded in a light weight container, the {\it BRPacket}. The use of this container 
ensures the control of time synchronization throughout the system and the different stages 
of processing thanks to an hardware 125 MHz clock time tag encoded in BRPacket.  
\item We have also developed the control and data acquisition software which run on a 
cluster of Linux-PC's. The control software uses the USB to configure DIGFFT boards and the serial 
RS232 interface for DCLK.  
The acquisition system is capable of sustained acquisition rates of several 
hundreds megabytes/second (MB/s) per PCI-Express receiver board. The software is object oriented, 
in C++ and multi-threaded. Its design ensures flexibility and very high efficiency. 
Data can be received through PCI-Express boards, or exchanged through the network interfaces. 
The software components can be assembled to perform several types of
\og on the fly processing\fg{},
distributed over 
the cluster computing nodes, taking advantage of the multi-cpu/multi-core processing power thanks to the 
multi-threaded software architecture.
In particular, the acquisition program can be configured as a high throughput software correlator 
which computes all pair wise visibilities for interferometric observations. 
\item A prototype for an FPGA based digital correlator / beam former is also being developed. FFT data from the 
DIGFFT boards is sent directly  to the beam former board. First tests of this beam former prototype should take place 
with the FAN prototype in Nan\c{c}ay in the next few months. 
\end{enumerate}
 
\begin{figure}
\centering
\mbox{
\vspace*{-15mm}
\includegraphics[width=0.7\textwidth]{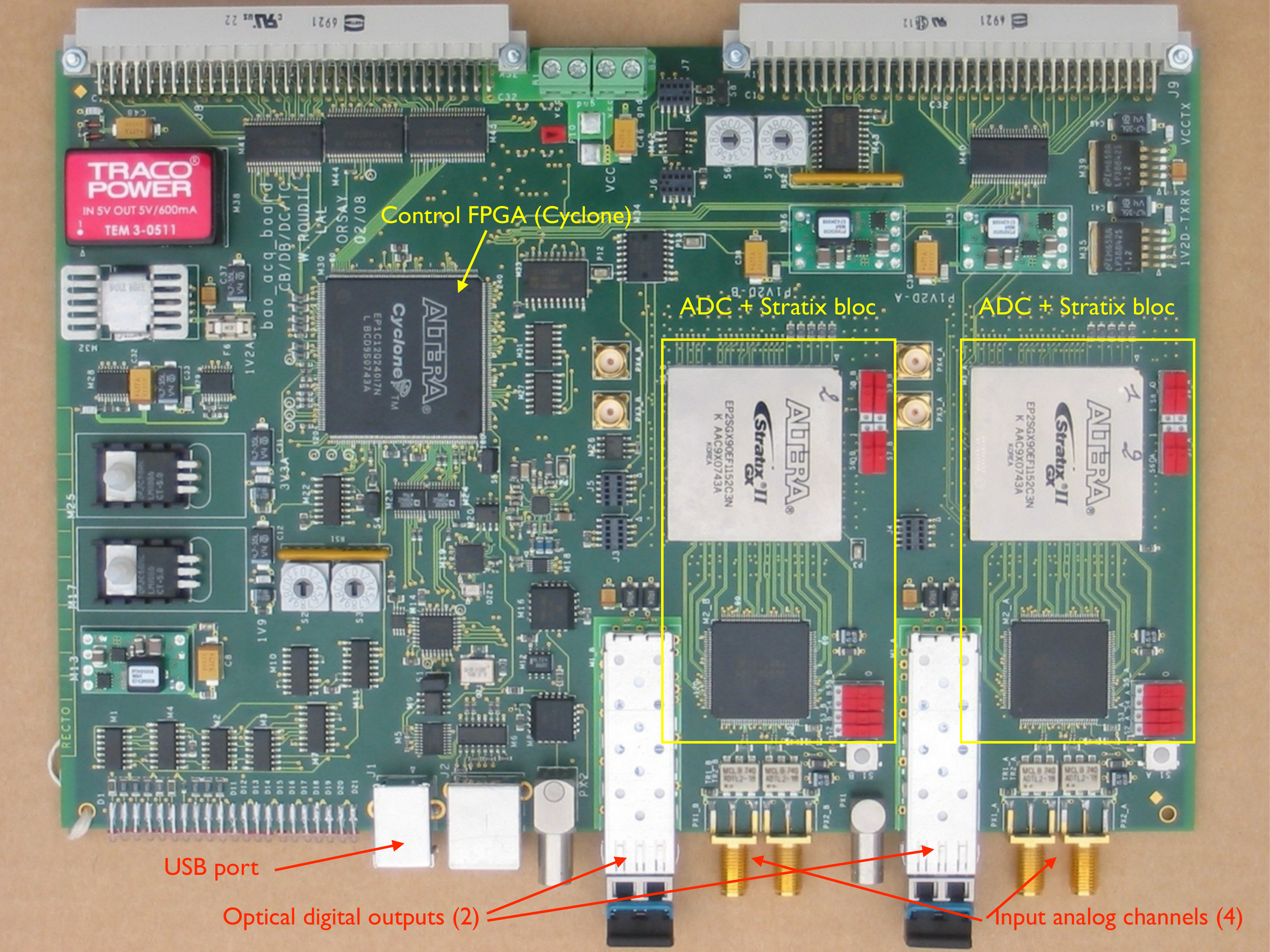}
}
\vspace*{0mm}
\caption{DIGFFT board (4 analog channels sampled at 500 MSample/s with 8 bits dynamic range) capable of digital filtering (FFT)  
developed for the BAORadio project.}
\label{figadcboard}
\end{figure}

\section{System qualification and performance}

The different subsystems have been tested using test benches in the laboratory, 
and the complete system has been qualified using the Nan\c{c}ay radio telescope (NRT). It has been used in interferometric 
mode on the CRT (Cylindrical Radio Telescope) prototype at CMU in Pittsburgh \cite{bandura.11}. 
The interferometric mode operation associated with the software correlator is also currently in use for 
beam forming with the FAN prototype. Major milestones for the system tests and qualification are listed here:
\begin{itemize} 
\item July 2008 : first test of a complete analog+digital electronic pipeline and the acquisition system with the NRT 
\item November 2009 : Observation campaign using the CRT prototype at Pittsburgh,  with 16 channels and the
FFT firmware. Visibilities were then computed offline, from FFT data saved to disk. 
\item November 2010 : Observation campaign using the CRT prototype at Pittsburgh, 32 channels, FFT on FPGA and software correlator.
\item 2011 : Observation and beam forming tests with the FAN prototype at Nan\c{c}ay, with 12 digitization channels; use of the BAORadio electronic system for the {\HI}-cluster program with NRT.
\end{itemize}

\begin{figure}
\centering
\vspace*{-15mm}
\mbox{
\includegraphics[width=0.85\textwidth]{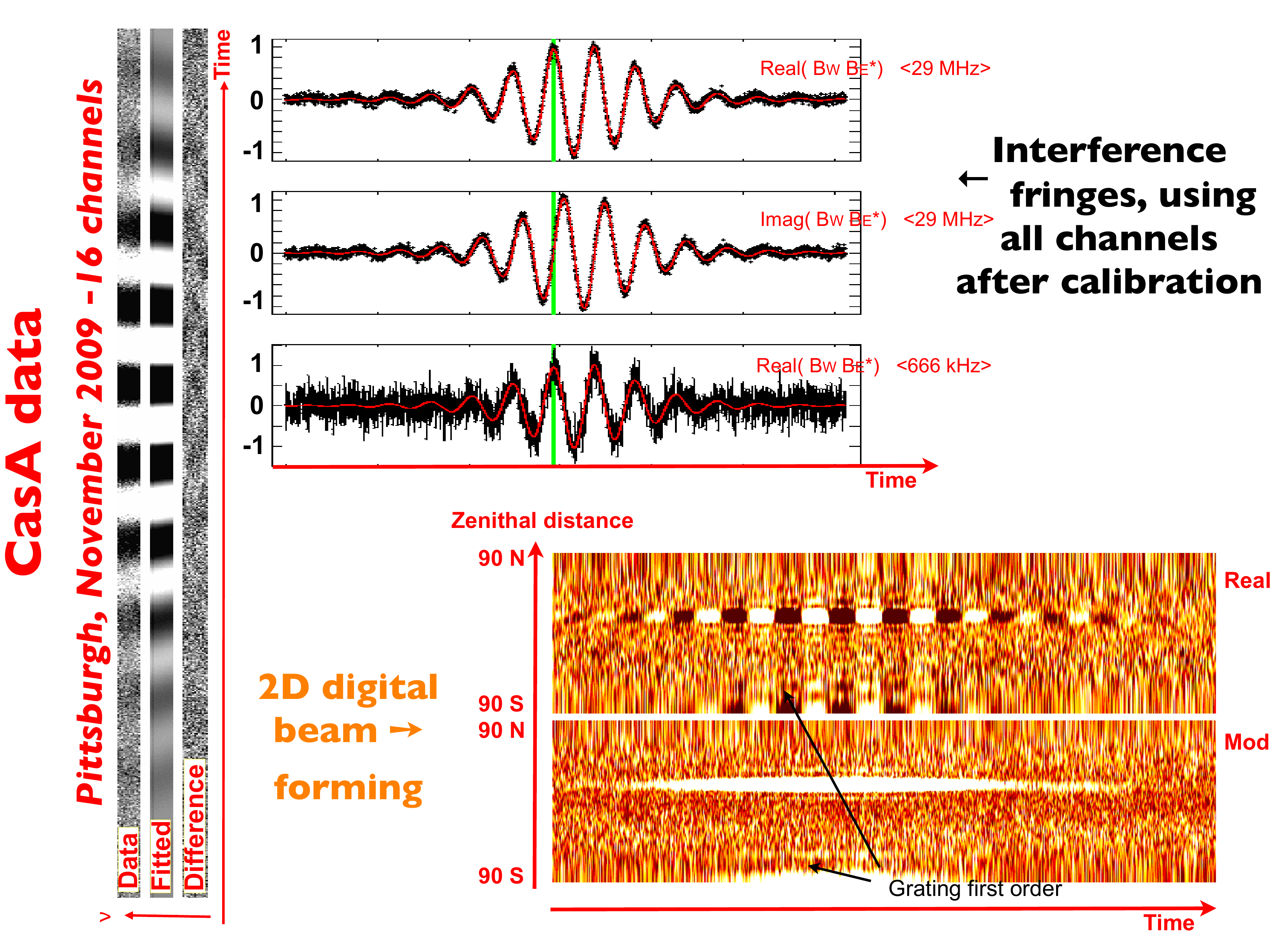}
}
\vspace*{-2mm}
\caption{Interference signal and digital beam synthesized; data taken
  in 2009 using Pittsburgh/CMU CRT prototype cylindrical during
  Cassiopeia A transit. 
The top diagram shows the interference fringes with the horizontal axis
representing a time span of about one hour.
The left bands show the time vs frequency interference signal for data
and for the fitted model, with the residue.
The lower diagram shows the digitally synthetized beam from the 16 receivers; 
the horizontal axis represents the time or right ascension, the
vertical axis corresponds to the zenithal distance in the meridian plane.}
\label{figpittsnov09}
\end{figure}

The CRT prototype at Pittsburgh is made of two cylindrical reflectors, each 25 meter long and 10 meter wide, 
equipped with half wavelength dipole receivers along the focal
line. The cylinder axis are oriented north-south 
and they are 25 meters apart. A small section of the two cylinders,
corresponding to 8 receivers along each 
cylinder, have been equipped with the BAORadio electronic system in November 2009. We have observed 
several bright radio source transits (Cassiopeia A, Cygnus A, Sun). Fourier coefficient data  computed by 
the FFT-firmware on the DIGFFT boards have been saved to disk by the acquisition system.
The 2009 observation campaign with the CRT at Pittsburgh enabled us to test the correct system operation, 
stability and time synchronization in interferometric mode, 
with several digitization streams (16 channels, 4 DIGFFT boards). 
We have also used these data to develop and debug 
the software correlator which has been subsequently used in Pittsburgh in 2010, and is being used for FAN beam 
forming. We have also developed  software for RFI cleaning, relative gain and phase calibration and 
digital beam forming using the computed visibility data. 
Figure \ref{figpittsnov09}  shows  examples of the results obtained after cleaning, calibrating and processing 
the November 2009 Pittsburgh data. The interference fringes between the two cylinders, using all 16 receivers
(8/cylinder), after gain and phase alignment are visible in the top part with very high signal to noise ratio, 
while the lower map shows the two dimensional, digitally synthesized
beam from the Cassiopeia A transit. 

The BAORadio system is currently used at Nan\c{c}ay for the $\mathrm{H_I -Cluster}$ program, which aims at the 
detection of the hydrogen gas in nearby clusters ($z_{max} \lesssim 0.12$) Abell 85, Abell 168, Abell 1225, Abell2440.
Since end of 2010, two orthogonal polarizations of the NRT cryogenic receiver are instrumented 
with the complete electronic chain described in section 3, in parallel 
with the standard readout system. Observations have been carried out nearly routinely over the last 
few months, and we have accumulated about 50 TeraBytes of data, corresponding to $\sim$ 50 hours observations, 
which are stored at the IN2P3 Computing facility in Lyon. The data analysis is under way, with the 
primary task being the data cleaning for RFI removal. 

\begin{figure}
\centering
\vspace*{-15mm}
\mbox{
\includegraphics[width=0.9\textwidth]{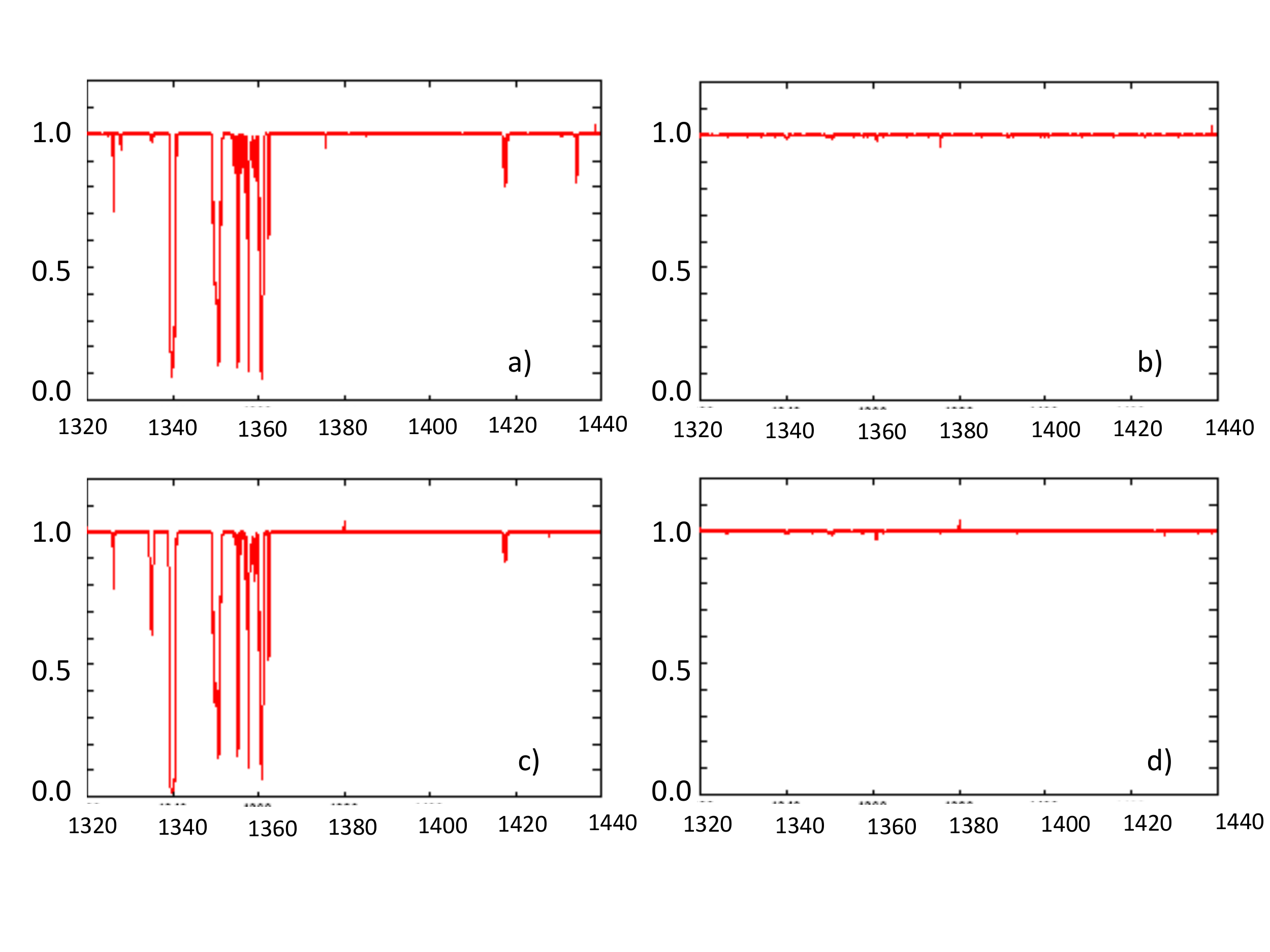}
}
\vspace*{-12mm}
\caption{Example of RFI suppression using a sample of 35 seconds of data taken for the 
$\mathrm{H_I -Cluster}$ program. Are displayed in the 1320-1440 MHz frequency band, 
the signal over noise ratios (S/N in arbitrary unit) for both
orthogonal polarisations (top and bottom panels)
without filtering (left panels) and with filtering (right panels).
In the case of intermittent RFI polluting the signal, S/N decreases
since N increases; in the case of a dominating constant RFI, S/N
increases since N decreases.
}
\label{figHICl}
\end{figure}

Each analog signal, corresponding to one of the 
two NRT polarizations, is sampled at 500 Msample/s, with $\sim$ 32 $\mathrm{\mu s}$ (16384 samples) 
digitization framelength, at a rate of 10000 frame/s. The data flow dumped to disk 
represents more than 300 MByte/s (about one TByte/hour) for the two channels, and corresponds to 
$\gtrsim 30 \%$ on sky efficiency. After offline Fourier transform (FFT), we obtain spectra 
with 30.5 kHz frequency resolution, covering the entire frequency band $[1250,1500] \mathrm{MHz}$ 
every 100 $\mathrm{\mu s}$ (10 kHz). The time-frequency data stream is then processed 
through frequency and time domain statistical filtering
to remove  RFI over the 250 MHz wide frequency band.
Figure \ref{figHICl} illustrates the results on RFI cleaning obtained by this data processing.
It can be seen that most of the spurious signals, due to terrestrial radio signals present in the 
frequency band 1320-1370 MHz has been removed. 
Application of similar methods to the data taken with the standard NRT correlator readout system 
has shown that RFI cleaning performance is enhanced by orders of magnitudes thanks to the
fine time sampling, combined with the frequency resolution and dynamic range of the BAORadio chain. 

\section{Conclusion}
3D mapping of the {\HI} mass distribution through 21 cm intensity mapping in radio, in the frequency range 
400-1500 MHz, is a promising method for cosmological surveys of matter distribution up to redshift $z \lesssim 3$,
and would allow one to constrain the dark energy equation of state parameters by measuring BAO scale evolution 
with redshift.
A multi-beam wide band instrument (few hundred MHz), FPA or interferometer, 
with a large field of view (20-100 deg$^2$) should be used to insure a
sufficiently large sensitivity for surveying 
a large volume of the universe at 21 cm. The electronic system presented here has been designed to equip 
multi-receiver systems, with few hundred digitization channels. The analog electronic modules (AEM), and
the sampling boards (DIGFFT) can be spread over an area of a few hectares. The digital signals are then 
transferred to a central facility through optical fibers. This central facility might use a custom build 
correlator/beam former, or a computer cluster with high bandwidth network. The computing 
nodes would be equipped with GPU or custom designed FPGA boards to handle the CPU intensive pair wise 
visibilities or beam forming computation. The electronic and
acquisition system has been successfully extensively 
tested and qualified at the Nan\c{c}ay radio observatory, as well as in Pittsburgh.



\vspace*{-5mm}
\section*{Acknowledgements}
\vspace*{-2mm}
We would like to thank the LAL and Irfu technical staff, 
P. Abbon, C. Beigbeder, D. Breton, T. Caceres, D. Charlet,, E. Delagnes, H. Deschamps, 
C. Flouzat, P. Kestener, B. Mansoux, C. Pailler, M. Taurigna, who have designed and built 
the electronic and acquisition system presented in this paper. We are also grateful to the 
Nan\c{c}ay station staff for their help, in particular J. Pezzani and C. Dumez-Viou 
for their expertise and advice during system qualification.

\end{document}